\documentclass[preprint,nofootinbib,amsmath,amssymb]{revtex4}

\usepackage[dvips]{graphicx}
\usepackage{amsmath,amsthm,amssymb}

\begin{document}
\title{The solar system mimics a hydrogen atom.}

\author{Je-An Gu}
\email{jagu@ntu.edu.tw}

\affiliation{Leung Center for Cosmology and Particle
Astrophysics, National Taiwan University, Taipei, 10617,
Taiwan}

\begin{abstract}
The solar system and the hydrogen atom are two well known
systems on different scales and look unrelated: The former is a
classical system on the scale of about billions of kilometers 
and the latter a quantum system of about tens of picometers. 
Here we show a connection between them. Specifically, we find
that the orbital radii of the planets mimic the mean radii of
the energy levels of a quantum system under
the Coulomb-like potential. 
This connection might be explained by very light dark matter
which manifests quantum behavior in the solar system, thereby
hinting at a dark matter mass around $8 \times 10^{-14}$
electron-volts.
\end{abstract}

\maketitle


\begin{figure}[h!]
\includegraphics[width=\textwidth]{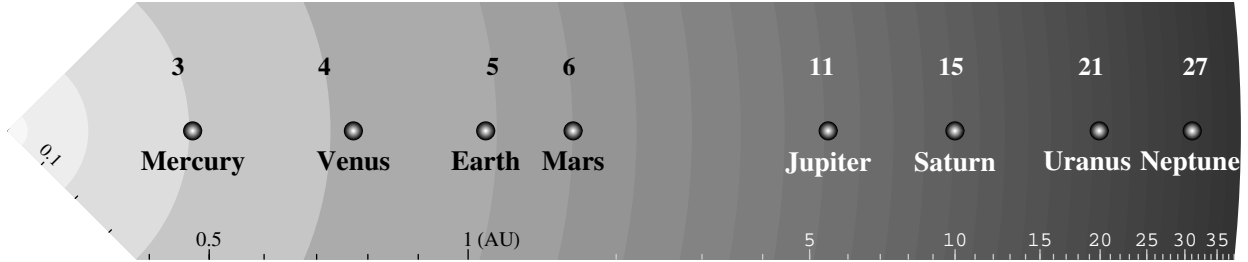}
\caption{\label{fig1:conformity} \small
The solar system mimics a hydrogen atom. The orbital radii of
the planets are shown in the middle row. The gray levels
illustrate the mean radii of thirty \textit{s}-state energy levels of a hydrogen
atom which has been enlarged about $8 \times 10^{19}$ times.
The eight numbers in the top row are the quantum numbers of the
energy levels which the planets conform to. The ticks and the
numbers at the bottom denote
the distance from the center (the tip on the left) 
in unit of AU; the distance below 1AU is on a linear scale and
that beyond 1AU on a log scale. ($1\textrm{AU} \simeq 1.5
\times 10^{11} \textrm{meters}.$)}
\end{figure}

For a quantum system under the Coulomb-like potential
$V(r)\propto -1/r$ such as a hydrogen atom,
the mean radius 
for an energy eigenstate with zero azimuthal quantum number
(\textit{s} state) is proportional to $n^2$, where the
principle quantum number $n=1,2,\ldots$ (e.g., see \cite{QM}).
Here we examine whether the solar system
\cite{NASA} has the same feature, regarding 
the orbital radii of the planets. Specifically, we attempt to
find smaller positive integers $n_i$
($i=1,2,\ldots,8$) 
such that $n_i^2 r_0$ are close to the orbital radii of the
planets, where the proportionality constant $r_0$ is the
ground-state radius of the quantum system.

\begin{table}[h]
\caption{\label{table:conformity} The conformity of the
planetary orbits with the quantum energy levels.}
\begin{tabular}{lcccccccc}
  \hline \hline
  Planet & Mercury & Venus & Earth & Mars & Jupiter & Saturn & Uranus & Neptune \\
  \hline
  Mass \footnotesize ($10^{24}$kg) & 0.330 & 4.87 & 5.97 & 0.642 & 1898 & 568 & 86.8 & 102 \\
  Orbital radius {\footnotesize ($10^6$km)}$^\ast$ & 57.9 & 108.2 & 149.6 & 227.9 & 778.6 & 1433.5 & 2872.5 & 4495.1 \\
  \hline
  Radius $n_i^2 r_0$ {\footnotesize ($10^6$km)}$^\dag$ 
   & 56.8 & 101.0 & 157.9 & 227.3 & 764.1 & 1420.9 & 2784.9 & 4603.7 \\
  Quantum number $n_i$ & 3 & 4 & 5 & 6 & 11 & 15 & 21 & 27 \\
  Ground-state radius & \multicolumn{8}{l}{\hspace{0.8em} $r_0 = 6.315 \times 10^6$km} \\
  \hline
  Fractional error \footnotesize ($\%$) & $-1.84$ & $-6.62$ & $5.53$ & $-0.245$ & $-1.86$ & $-0.880$ & $-3.05$ & $2.41$ \\
  \multicolumn{9}{l}{Root-mean-square of the eight fractional errors: $\; 3.49\%$} \\
  \hline \hline
  \multicolumn{9}{l}{\footnotesize $^\ast$ The semi-major axis, i.e., the average distance from a planet to the sun.\vspace{-1em}}\\
  \multicolumn{9}{l}{\footnotesize $^\dag$ The expectation value of the radius for an \textit{s}-state energy level.}
\end{tabular}
\end{table}

The conformity 
of the solar system with a hydrogenlike quantum system is
depicted in Fig.~\ref{fig1:conformity}. The corresponding
quantum numbers $n_i$ of the planets are presented in both
Fig.~\ref{fig1:conformity} and Table~\ref{table:conformity}. As
shown in the table, the orbital radii of the planets are close
to the mean radii $n_i^2 r_0$ for the \textit{s}-state energy
levels. The fractional errors between them are several percent
or smaller; the root-mean-square (rms) of the eight fractional
errors is merely $3.49\%$. (Henceforth the rms of the
fractional errors will simply be termed the \textit{error}.)

The conformity with the lower energy levels is special 
and might be a sign of some fundamental physics behind. Note
that although the excellent conformity with very large $n_i$
can easily occur---the error can go to zero when $n_i$ go to
infinity---, however, the conformity we found with smaller
$n_i$ is unusual. We will show its rareness compared to the
cases of randomly generated orbital radii. We generate 10,000
sets of eight random radii; for each set we find the best-fit
of the positive integers $n_i$ under some low-energy-level condition (to be specified later) 
and obtain the error. With the distribution of the 10,000
errors we show the smallness of the error in the real case
compared to those in the random cases, e.g., via its deviation
from the median of the 10,000 errors. 

\begin{figure}
\includegraphics[width=0.6\textwidth]{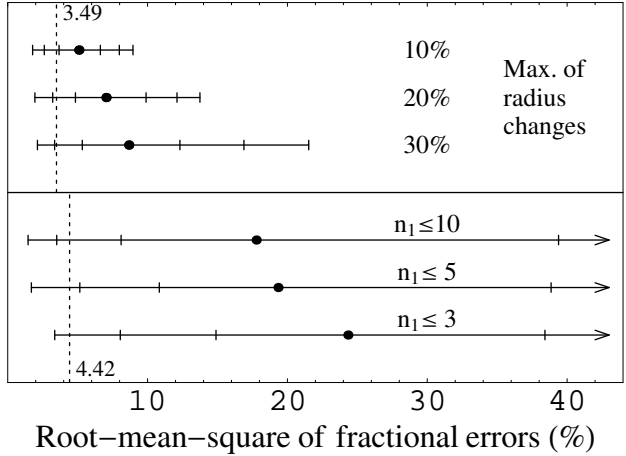}
\caption{\label{fig2:CI} \small %
The 1$\sigma$--3$\sigma$ confidence intervals for the median
(black dot) of the errors in the randomly generated cases. The
upper panel is for the three cases where the random radii are
generated within a certain fractional change of the real radii.
The lower panel is for those where the random radii of four
inner planets are fitted by four energy levels in a row under
some condition on $n_1$. The dashed lines denote the errors in
the real case: $3.49\%$ and $4.42\%$ respectively.}
\end{figure}

\begin{figure}
\includegraphics[width=0.6\textwidth]{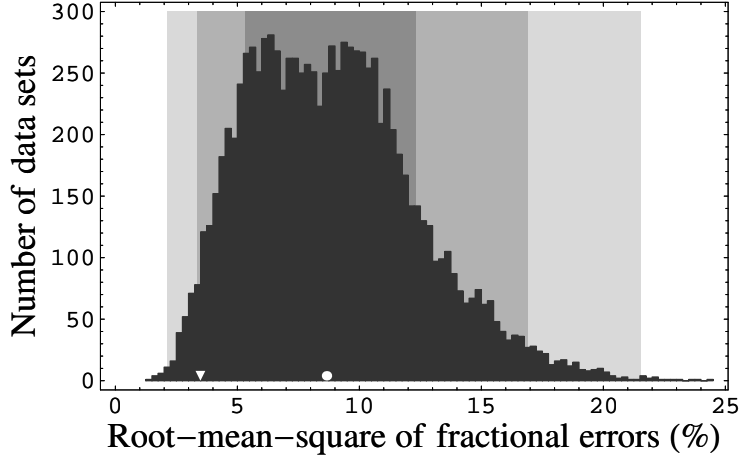}
\caption{\label{fig3:histogram1} \small %
The histogram of the errors in the third case of the upper
panel in Fig.~\ref{fig2:CI}. The median ($8.65\%$) of the errors
is denoted by the white dot, the real-case error ($3.49\%$) by
the white triangle, and the $1\sigma$--$3\sigma$ confidence
intervals for the median by the gray regions.}
\end{figure}

Two kinds of random radii are considered here. First, we
attempt to see how the goodness of conformity changes when the
planets are relocated to other orbits slightly different from
the real ones, with the fractional changes of radii within
$10\%$, $20\%$, or $30\%$.
For each case, we generate 10,000 sets of random radii with a
uniform probability distribution in the required range, and
compare the real case with them in the way described above. To
concentrate on the conformity with the lower energy levels, we
impose simply the condition $n_1 \leq 3$ but none for other
$n_i$ when finding the best-fit of $n_i$. The results are shown
in the upper panel of Fig.~\ref{fig2:CI}, which plots the
$1\sigma$--$3\sigma$ confidence intervals for the median of the
10,000 errors as well as the real-case error ($3.49\%$) for
comparison. The error in the real case is significantly smaller
than the median of the random ones, with the deviation
as follows: 
\begin{center}
\begin{tabular}{cc}
  \hline \hline
  Changes \hspace{1em} & Deviation \vspace{-0.6em} \\
  of orbital radii \hspace{1em} & from the median \\
  \hline
  $< 10\%$ \hspace{1em} & $1.17\sigma$ ($75.7\%$CL) \\
  $< 20\%$ \hspace{1em} & $1.81\sigma$ ($92.9\%$CL) \\
  $< 30\%$ \hspace{1em} & $1.92\sigma$ ($94.5\%$CL) \\
  \hline \hline
  \multicolumn{2}{c}{\footnotesize Fitting condition: $n_1 \leq 3$.}
\end{tabular}
\end{center}
The deviation increases with the range of the radius change. It
is near to $2\sigma$ in the $30\%$ case. For this case,
Fig.~\ref{fig3:histogram1} plots the distribution of the 10,000
errors together with the confidence intervals (gray regions),
the median (white dot) and the real-case error (white
triangle). These results show that the conformity of the solar
system with a hydrogenlike quantum system is special: A change
of the orbital radii of the planets will more likely reduce the
conformity.

\begin{figure}
\includegraphics[width=0.6\textwidth]{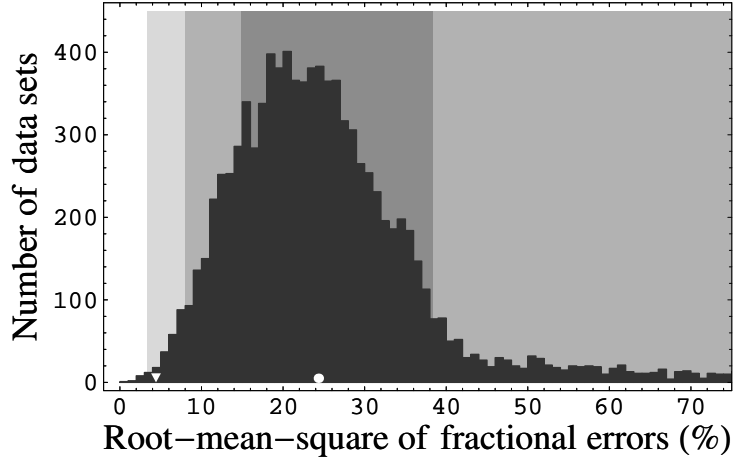}
\caption{\label{fig4:histogram1} \small %
The histogram of the errors in the third case of the lower
panel in Fig.~\ref{fig2:CI}. The median ($24.3\%$) of the errors
is denoted by the white dot, the real-case error ($4.42\%$) by
the white triangle, and the $1\sigma$--$3\sigma$ confidence
intervals for the median by the gray regions.}
\end{figure}

Second, we concentrate on the inner planets that are in
conformity with the energy levels $n_{1,2,3,4}=3,4,5,6$. %
Note that having the conformity of the inner planets with the
lower levels is more difficult, therefore more significative,
than that of the outer planets with the higher levels.
Moreover, as to be shown, the conformity with four levels in a
row is truly exceptional. 
We generate 10,000 sets of four positive random numbers with a
uniform probability.\footnote{An upper bound to the random
numbers is required. However, its value is irrelevant here
because it does not change the error but simply changes the
proportionality constant $r_0$.} For each set, we obtain the
best-fit of $n_i$ under the conditions that $n_{1,2,3,4}$ are
four positive integers in a row and that $n_1 \leq 10$, $5$, or
$3$. The results are shown in the lower panel of
Fig.~\ref{fig2:CI}. The smallness of the real-case error
($4.42\%$) compared to the random ones is even more significant
than the previous cases,
with the deviation from the median as follows: 
\begin{center}
\begin{tabular}{lrl}
  \hline \hline
  $n_1\leq   10$   & : & \hspace{1em} $1.76\sigma$ ($92.2\%$CL) \\
  $n_1\leq \, 5$  & : & \hspace{1em} $2.17\sigma$ ($97.0\%$CL) \\
  $n_1\leq \, 3$  & : & \hspace{1em} $2.79\sigma$ ($99.5\%$CL) \\
  \hline \hline
\end{tabular}
\end{center}
As expected, the deviation 
gets larger when the condition on $n_i$ gets tighter. Even in
the case with a rather loose condition, $n_1 \leq 10$, the
real-case error has already sat outside the $90\%$ confidence
interval.
It further goes beyond the $99\%$ confidence interval 
in the case where $n_1 \leq 3$. For this case,
Fig.~\ref{fig4:histogram1} plots the distribution of the 10,000
errors together with the confidence intervals, the median 
and the real-case error. 
These results show that it is very rare to have the conformity
with four energy levels in a row as good as that of the inner
planets of the solar system.

The conformity might be explained by very light dark matter. In
cosmology it is widely believed that dark matter helps the
formation of the cosmic structures such as galaxies, galaxy
clusters etc: The dark matter structures formed beforehand and
baryons followed later; i.e., after the recombination of
electrons and protons, baryons fell into the gravitational
potentials provided by the dark structures.
Here we 
speculate the possible role of dark matter in the formation of
the solar system 
and give a sketch of the scenario. Specifically, we consider
the dark matter which is so light that its de Broglie
wavelength is on the scale of the solar system and therefore
its mass distribution in the solar system
manifests quantum behavior.
\footnote{For even lighter dark matter with its quantum nature
manifest at galactic scales or beyond, see, e.g.,
\cite{ultralightDM}.} The mass distribution of the dark matter
structure formed beforehand may respect the wave functions of
the energy eigenstates. Since the radial probability
distribution of an energy eigenstate is peaked roughly around
its mean radius, with a smaller width for a lower energy level,
the mass density of dark matter is larger around these mean
radii of the energy levels.
Later, when the embryos of the planets formed via the accretion
of dust grains, the denser regions of dark matter may provide
nuclei for the accretion; thereby the planetary orbits can be
related to the quantum energy levels of dark matter.

The formation of the planets guided by different energy levels
may have different fates: \vspace{-0.3em}
\begin{enumerate}
  \item The first few levels are too close to the sun to form planets. \vspace{-0.6em} %
  \item Each of the next several lower levels helps to form an inner planet. \vspace{-0.6em} %
  \item Several of the higher levels together help to form an outer, more massive planet, %
        encircled by a ring system and orbited by many moons. \vspace{-0.3em} %
\end{enumerate}
These three situations are exhibited
in Table \ref{table:conformity}: no planet at $n=1,2$, one
planet at each of $n=3,4,5,6$, and one planet every 4--6 higher levels. 
Note that Fate 3 for the higher levels is possibly due to the
wide spread of the probability (therefore the mass)
distribution as well as the small energy difference between the
nearby levels that makes the transition between them easy to
occur.

To estimate the mass of dark matter, we consider dark matter
under the central potential $V(r) = -G M_\odot m / r$, where
$M_\odot$ is the mass of the sun and $m$ the dark matter mass.
Under this potential the mean radius for an \textit{s}-state
energy level is $(3/2)n^2 a_0$, where $a_0 \equiv (G M_\odot
m^2)^{-1}$ (analogous to the Bohr radius), and accordingly the
ground-state radius $r_0 = (2 G M_\odot m^2 / 3)^{-1}$.
Considering $n_i$ and $r_0$ in Table \ref{table:conformity} as
the best-fit with the \textit{s}-state energy levels, 
we obtain the dark matter mass $m \simeq 8 \times 10^{-14} $ 
electron-volts.

The conformity of the solar system with a quantum system
indicates the quantum nature of the solar system, which is
possibly carried by very light dark
matter. 
It suggests the possibility that dark matter or some other
quantum source plays an important role in the formation of the
planets. It invites the study of the formation and the
evolution of the solar system with dark matter taken into
consideration. This may give a different story of the solar
system. It is also worth investigating the exoplanet systems,
e.g., examining the conformity of their orbits with the energy
levels of a quantum system and, if explained by dark matter,
evaluating its mass. If many planet systems exhibit this
conformity and suggest similar mass of dark matter, they will
give a strong support to this scenario of the planet formation
and to very light dark matter, and can serve as a new probe of
dark matter.


\begin{thebibliography}{0}

\bibitem{QM}
  R.~Eisberg and R.~Resnick,
  ``Quantum Physics of Atoms, Molecules, Solids, Nuclei, and Particles,'' 
  John Wiley $\&$ Sons (1985).

\bibitem{NASA}
  NASA, ``Planetary Fact Sheet,'' http://nssdc.gsfc.nasa.gov/planetary/factsheet/

\bibitem{ultralightDM}
  W.~H.~Press, B.~S.~Ryden and D.~N.~Spergel,
  ``Single Mechanism for Generating Large Scale Structure and Providing Dark Missing Matter,''
  Phys.\ Rev.\ Lett.\  {\bf 64}, 1084 (1990);\\
  S.-J.~Sin,
  ``Late time cosmological phase transition and galactic halo as Bose liquid,''
  Phys.\ Rev.\ D {\bf 50}, 3650 (1994)  [hep-ph/9205208];\\
  V.~Sahni and L.-M.~Wang,
  ``A New cosmological model of quintessence and dark matter,''
  Phys.\ Rev.\ D {\bf 62}, 103517 (2000)  [astro-ph/9910097];\\
  W.~Hu, R.~Barkana and A.~Gruzinov,
  ``Cold and fuzzy dark matter,''
  Phys.\ Rev.\ Lett.\  {\bf 85}, 1158 (2000)  [astro-ph/0003365].

\end{thebibliography}
\end{document}